\documentclass[aps,twocolumn,floatfix,superscriptaddress]{revtex4}

\usepackage[normalem]{ulem}

\usepackage{flushend}

\usepackage{fancyhdr}
\usepackage{kantlipsum}

\usepackage{amsmath}
\usepackage{graphicx}

\usepackage{upgreek}

\usepackage{color}

\usepackage{floatrow}

\usepackage[unicode=true,bookmarks=false,breaklinks=false,pdfborder={0  0 1},backref=false,colorlinks=false]{hyperref} 
\usepackage{breakurl}

\usepackage{lettrine}

\setcitestyle{round}

\makeatletter

\renewcommand{\figurename}{\textbf{Fig.}}
\renewcommand{\thefigure}{\textbf{\arabic{figure}}}

\@ifundefined{textcolor}{}
{
 \definecolor{BLACK}{gray}{0}
 \definecolor{WHITE}{gray}{1}
 \definecolor{RED}{rgb}{0.7,0,0}
 \definecolor{GREEN}{rgb}{0,1,0}
 \definecolor{BLUE}{rgb}{0,0,1}
 \definecolor{azul}{rgb}{0.8,0.85,0.9}
 \definecolor{text}{rgb}{0.1,0.4,0.65}
 \definecolor{CYAN}{cmyk}{1,0,0,0}
 \definecolor{MAGENTA}{cmyk}{0,1,0,0}
}

\newcommand{\nanotec}{
CNR NANOTEC, Istituto di Nanotecnologia, 73100 Lecce, Italy
}

\newcommand{\lecce}{
Istituto Italiano di Tecnologia, Center for Biomolecular Nanotechnologies Lecce, 73010
Arnesano, Italy
}

\newcommand{\warwick}{
Department of Physics, University of Warwick, CV4 7AL Coventry, United Kingdom
}

\newcommand{\mexico}{
Instituto de Energ\'{i}as Renovables, Universidad Nacional Aut\'{o}noma de M\'{e}xico, Temixco, Morelos 62580, Mexico
}

\newcommand{\lkb}{
Laboratoire Kastler Brossel, Pierre and Marie Curie University,
Sorbonne Universit\'{e}s, CNRS, 
\'{E}cole Normale Sup\'{e}rieure, Paris Sciences et Lettres Research University,
Coll\`{e}ge de France, F-75005 Paris, France
}

\newcommand{\ucl}{
Department of Physics and Astronomy, University College London, WC1E 6BT London, United Kingdom
}

\begin{document}

\title{
{
\usefont{OT1}{cmss}{m}{n}
{\LARGE 
\textbf{
Twist of generalized skyrmions and spin vortices \\ in a polariton superfluid
}
}
}
}


\author{Stefano Donati}
\affiliation{\nanotec}
\affiliation{\lecce}

\author{Lorenzo Dominici}
\email{lorenzo.dominici@gmail.com}
\affiliation{\nanotec}

\author{Galbadrakh Dagvadorj}
\affiliation{\warwick}

\author{Dario Ballarini}
\affiliation{\nanotec}

\author{Milena De Giorgi}
\affiliation{\nanotec}

\author{Alberto Bramati}
\affiliation{\lkb}

\author{Giuseppe Gigli}
\affiliation{\nanotec}

\author{Yuri G.~Rubo}
\affiliation{\mexico}

\author{Marzena Hanna Szyma\'{n}ska}
\email{m.szymanska@ucl.ac.uk}
\affiliation{\ucl}

\author{Daniele Sanvitto}
\affiliation{\nanotec}

\begin{abstract}
{

}
\end{abstract}

\maketitle

\noindent\textbf{We study the spin vortices and skyrmions
coherently imprinted into an exciton-polariton condensate on a planar
semiconductor microcavity. We demonstrate that the presence of
a polarization anisotropy can induce a complex dynamics of these structured topologies, 
leading to the twist of their circuitation on the Poincar\'{e}
sphere of polarizations. The theoretical description of the results carries the
concept of generalized quantum vortices in two-component superfluids, which
are conformal with polarization loops around an arbitrary
axis in the pseudospin space.}

\vspace{0.3cm}

\noindent{\small
quantum vortices | topology | Poincar\'{e} space | \\ condensates | polaritons
}

\vspace{0.3cm}

\usefont{OT1}{ppl}{m}{n}

\lettrine{T}opological 
defects represent a wide class of objects relevant to different fields
of physics from condensed matter to cosmology. The universality of
monopoles, vortices, skyrmions, domain walls, and of their
  formation processes in different systems, has largely motivated
  their study in the condensed matter context. 
  In particular, the interplay between the symmetry breaking in phase transitions and the formation of topological defects has been the focus of intensive research in the last century. 
In high energy physics,
the existence of an isolated point source
intrigued a great number of
physicists~\cite{Shnir2005}. Dirac was ``surprised if
Nature had made no use of it'' and postulated the
  possibility of the magnetic monopoles linked to the quantization of
electric charge~\cite{Dirac1931}. However the elusiveness of their
observation in free space has motivated an extensive study
of monopole analogues in the form of quasiparticles in many-body
systems~\cite{Volovik2003} such as the exotic spin ices~\cite{Castelnovo2008,Morris2009},
liquid crystals~\cite{Chuang1991}, exciton-polariton~\cite{Hivet2012}
and rubidium Bose-Einstein condensates (BECs)~\cite{Ray2014, Ray2015},
as well as other systems~\cite{Fang2003, Milde2013}. In a 2D
multicomponent BEC, an equivalent topological structure to the monopole is given by
the hedgehog polarization vortex~\cite{Toledo2014} which together with the hyperspin
vortex~\cite{Manni2013} belong to the class of spin vortices than when combined lead to
a well defined polarization pattern. 
Such topological states are characterized by a linear polarization vector which rotates an integer number of times (spin winding
number) around a singular central point, in a way analogue to what the magnetization does in the spin
vortices of a ferromagnetic spinor BEC~\cite{Sadler2006}.

In this work we excite complex vortex states in an 
exciton-polariton superfluid and study their amplitude, phase and polarization dynamics. 
We demonstrate that temporal evolution of such topologies leads, in general, 
to a twist of the polarization plane in the Poincar\'{e} space. 
The observed features of the
vortex behavior are explained within the concept of generalized
spin vortices, where the rotation of polarization at large
distances occurs around an arbitrary axis on the Poincar\'{e}
sphere.

Polaritons emerge in planar semiconductor microcavities as eigenmodes
of the strong coupling regime between the exciton resonance and the
photon cavity mode, combining the properties of light and matter. The
photons confer on polaritons a very small effective mass
($10^{-4}-10^{-5}$ of free-electron mass) which, together with
nonlinear interactions due to the excitons, leads to effective
condensation at a relatively high temperature (up to room temperature
for given materials such as ZnO, GaN or organic dyes).  Features
related to superfluidity have been observed such as the suppression of
scattering from defects (zero viscosity)~\cite{Amo2009a,Amo2009b} or
persistence of vortex currents~\cite{Sanvitto2010}.  Moreover,
polaritons with the pseudospin, given by the possibility of polarizing their state,
open the opportunity to study condensates with an internal angular
momentum degree of freedom, easily detected by optical means thanks to
their photonic outcoupling features~\cite{Shelykh2010}.

In spinor superfluids, the spin degrees of freedom allow for different composite topologies which emerge as the superposition of quantized vortex states. In superfluids with unrestricted geometry, these elementary vortex blocks are half-quantum vortices (HQVs)~\cite{Volovik2003}. In exciton-polariton condensates, the HQV is characterized by a $\pi$ phase rotation accompanied by a $\pi$ linear polarization rotation in such an elegant way that the two combine together to ensure the global continuity of the spinor wave function~\cite{Rubo2007,Lagoudakis2009,Manni2012,Dominici2015}. For finite-size condensates, where the boundary condition at large distances is not fixed, a HQV can be transformed into a skyrmion. The skyrmions possess a specific circumference of full linear polarization and they are fingerprinted by the inversion of the sign of circular polarization degree when crossing this circumference in the radial direction.
  
On the other hand, the dynamics of the pseudospin vector in semiconductor microcavities is
related to the presence of spin-orbital-like coupling, namely,
the transverse-electric--transverse-magnetic (TE-TM) splitting of the
modes, which is manifested by the optical spin Hall effect~\cite{Kavokin2005,Leyder2007,Kammann2012}.
 The TE-TM splitting is often represented by means of an
effective magnetic field that produces a precession of the pseudospin
vector, which leads to different sectors in circularly
polarized states both for real and momentum space, even when starting with a
homogeneously polarized
field~\cite{cilibrizzi_skyrmion_2016,Cilibrizzi2015}. 

Here we are able to initialize the polariton condensate 
with non-trivial pseudospin patterns.
We study the dynamics of exotic topologies such as the lemon and the star skyrmion, the hedgehog and the hyperspin vortex,
in clean regions of the sample, 
deriving universal observations not linked to
specific local disorder/defects
pinning~\cite{Manni2013}
or to the effect of sample architecture/confinement~\cite{Dufferwiel2015}.
The resultant topologies allow to extend the concept of quantum vortices into a wider class which includes states as generalized skyrmions and spin vortices.
These observations subtend the potentialities of resonant excitation of spin and orbital angular momentum states on microcavity polariton fluids,
and of their full control using TE-TM or anisotropy splitting, 
which is of fundamental importance
in the field of spintronics and polarization shaping.\\ 

\noindent \textbf{Setting up skyrmions and spin vortices\\}
Advanced phase-shaping was recently obtained by means of anisotropic
and inhomogeneous liquid-crystal devices called
$q$-plates~\cite{Cardano2012}, which allows an extensive
investigation of optical vorticity
and of full- and half-quantum vortex dynamics in polariton
condensates~\cite{Dominici2015}. 
Phase-shaping is applied to an initial Laguerre-Gauss $LG_{00}$
laser pulse (4 ps duration and 0.5 nm bandwidth) by sending it
across the $q$-plate carrying unitary topological charge. 
The $LG_{00}$ is hence partially or completely transformed into a unitary
winding state breaking the chiral symmetry between the two spin
populations. Upon proper setting of the incoming/outgoing polarization
and tuning of the $q$-plate, we can prepare the specific combination
and resultant field pattern (see \textcolor{blue}{\hyperref[sec:Sinfo]{\textit{Supporting Information}}}
 for experimental details).  
Indeed, each skyrmion and spin
vortex shown here can be thought as a composite state resulting from
the specific superposition of two $LG$ beams with integer
phase winding ($LG_{0,-1/0/+1}$), one in each of the two spin
components.

\begin{figure}[htbp]
\centering
\includegraphics[width=1.00\linewidth]{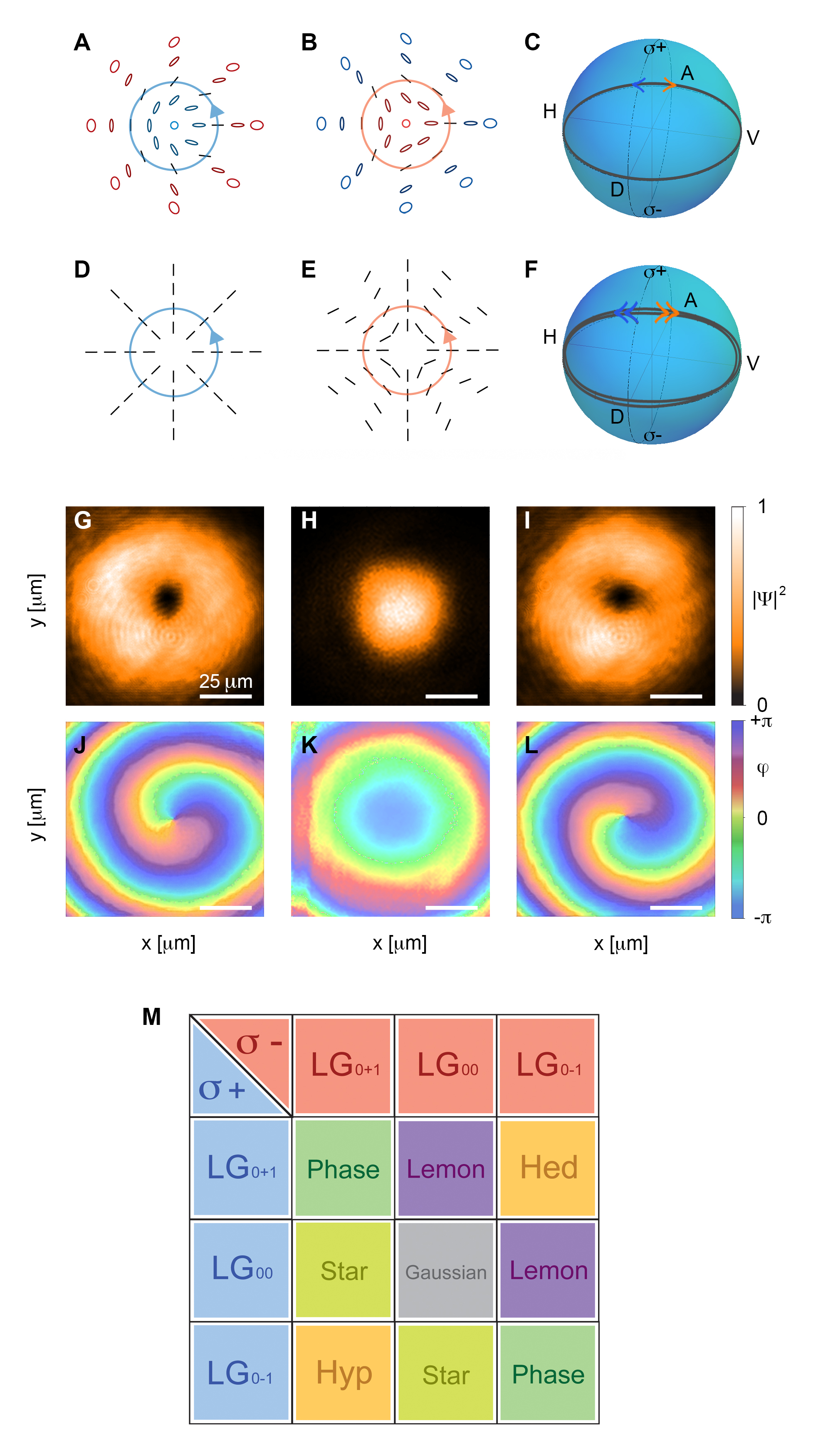}
\caption {Generation of vortices.
   Polarization fields map in real space relating to skyrmion (\textit{A},\textit{B}) and
   spin (\textit{D},\textit{E}) vortices. The skyrmion cover all the
   polarization states (full Poincar\'{e} patterns), giving rise to
   lemon-like (\textit{A}) and star-like (\textit{B}) patterns. Hedgehog (\textit{D}) and
   hyperspin (\textit{E}) feature a linear polarization in the whole space,
   following the radial direction and a hyperbolic pattern,
   respectively. The circles in the real space maps (which represent the $l$-line in the case of the skyrmions) 
   correspond to the
   equatorial line in the polarization spheres, with the associated
   direction for the different states. 
The colors in the polarization field maps, blue and red, are associated to either the $\sigma_+$ ($R$, right) or $\sigma_-$ ($L$, left) 
degree of circular polarization, respectively, while black is associated to the linear polarization.
(\textit{C},\textit{F}) Conformal mapping onto
   the Poincar\'{e} sphere of the real-space circle line around the
   vortex cores. Spin vortices follow a double rotation in
   the Poincar\'{e} space, skyrmions a single one. (\textit{G-I})
   Experimental initial Laguerre-Gauss states of the polariton population
   with (\textit{J-L}) the associated phase maps. Respectively, $LG_{0-1}$ (\textit{G},\textit{J}),
   $LG_{00}$ (\textit{H},\textit{K}) and $LG_{0+1}$ (\textit{I},\textit{L}). (\textit{M}) The table summarizes
   all the possible combinations of $LGs$ in the two cross spin
   polarizations of the system leading to the different states of quantum
   vortices: phase vortices (also known as full-vortices), hedgehog (Hed), hyperspin (Hyp), lemon
   and star skyrmions.} 
\label{fig:FIG1} 
\end{figure}

Projected onto the circular polarization basis, the skyrmions are
characterized by the presence of an integer phase winding (orbital
angular momentum) in one of the spin components and a zero-winding in
the opposite one. 
The resultant vectorial field exhibits an
inhomogeneous pattern comprising all polarization states, as typical
of full Poincar\'{e} beams. There exists a circle line in
real-space featuring linear polarization states ($l$-line, at $r=r_l$), which maps
to the equatorial loop of the Poincar\'{e} sphere (Fig.~1\textit{C}). 
The points inside or outside of the circle
are associated with either one or the other hemisphere of the sphere.
According to the skyrmion definition,
the pseudospin vector flips from right-circular at the core to left-circular (or viceversa) at its boundary ($r\approx 2r_l$).
Hence, the real-space radius maps to a given meridian on the sphere,
and the meridian angle is then associated to the azimuthal real-space angle.
The skyrmion polarization field in the region $r<r_l$ covers only one hemisphere of the Poincar\'{e} sphere,
 and it can be mapped to the polarization field of an infinite-size HQV~\cite{Rubo2007}. 
Therefore, similarly to the HQVs~\cite{Toledo2014}, 
the skyrmions can be characterized by two distinct geometries: lemon-like (Fig.~1\textit{A}) and star-like (Fig.~1\textit{B}) skyrmions.

In a spin vortex, the two spin components feature
counter-rotating phase windings (see
Fig.~1\textit{D},\textit{E})~\cite{Manni2013}. 
The central phase singularities in the two spin populations 
convert into a polarization singularity at the core.
There are two principal types of polarization
vortices: the hedgehog with a purely radial direction of
polarization (Fig.~1\textit{D}) and the hyperspin vortex, characterized by a
hyperbolic polarization pattern (Fig.~1\textit{E}).  
Upon changing the phase delay between the two spins, 
the hedgehog can transform into an azimuthal polarization pattern,
while the hyperspin undergoes a texture rotation.
These vortex states can be
described by an equivalent form when conformally mapping them to
points on the Poincar\'{e} sphere: the circulation along every circle
centered at the vortex core in the real-space can be associated
with a closed double loop lying in the equatorial plane (Fig.~1\textit{F}) of
the pseudospin space.
In order to make the classification clearer, in the third and
fourth rows of Fig.~1 we show examples of the density and phase
profiles of the fundamental building blocks of all type of
  vortices in circular polarization basis: clockwise $LG_{0-1}$
(\textit{G},\textit{J}), zero-winding $LG_{00}$ (\textit{H},\textit{K}) and counter-clockwise $LG_{0+1}$
(\textit{I},\textit{L}) states. Different combinations of the three $LG\text{s}$ in the
  two circular polarizations $\sigma_+$ and $\sigma_-$ (the two
  opposite pseudospins) leading to skyrmion, spin and phase
  vortices are shown in Fig.~1\textit{M}.

These photonic states are set as the initial conditions of
the polaritonic population dynamics by resonant excitation on
the microcavity sample, at the energy of the lower polariton branch
(LPB). On the detection side, we extract both the instantaneous local
density and the phase of the polariton emission during its time evolution,
by means of an interferometric setup performing real-time digital
Fourier transform and off-axis selection~\cite{Dominici2014,Anton2012,Nardin2010}. 
Using polarization filtering in detection, we can project
the single components in each of the three polarization basis, the
linear horizontal-vertical ($H$-$V$), diagonal-antidiagonal ($D$-$A$)
and the circular right-left ($R$-$L$). From these measurements it is
possible to retrieve the full map of the different degrees of
polarization, and Stokes parameters $S_{1,2,3}$, respectively, which
are used to plot the resultant polarization vector field, and to
associate every point in real-space to the pseudospin
space~\cite{Colas2015} (see \textcolor{blue}{\hyperref[sec:Sinfo]{\textit{Supporting Information}}}).\\

\noindent \textbf{Twist of the vortex polarization field\\}
The variation of the $S_{1,2,3}$ parameters in real-space for the
  hyperspin polarization vortex is presented in the first row of
  Fig.~2 (\textit{A-C}). These plots relate to the emission from the
polariton condensate at the initial time (after the laser pulse has
arrived), and the degrees of polarizations are clearly mapped also in
the regions of weak or null intensity, such as in the centre of a
vortex. The twofold symmetry ($C_2$ symmetry) of the $S_{1}$ (Fig.~2\textit{A})
and $S_{2}$ (Fig.~2\textit{B}) results in a petal shape of the polarization
distribution, while $S_{3}$ is approximately zero over all space as expected. The
presence of a very small component in the $S_3$ can be attributed to
the $q$-plate device, which cannot be simultaneously tuned at all the
wavelengths composing the pulsed beam. However, the circular degree of
polarization is far weaker than the linear components of the $S_{1}$ and
$S_{2}$ parameters.

\begin{figure}[htbp]
\centering
\includegraphics[width=1.0\linewidth]{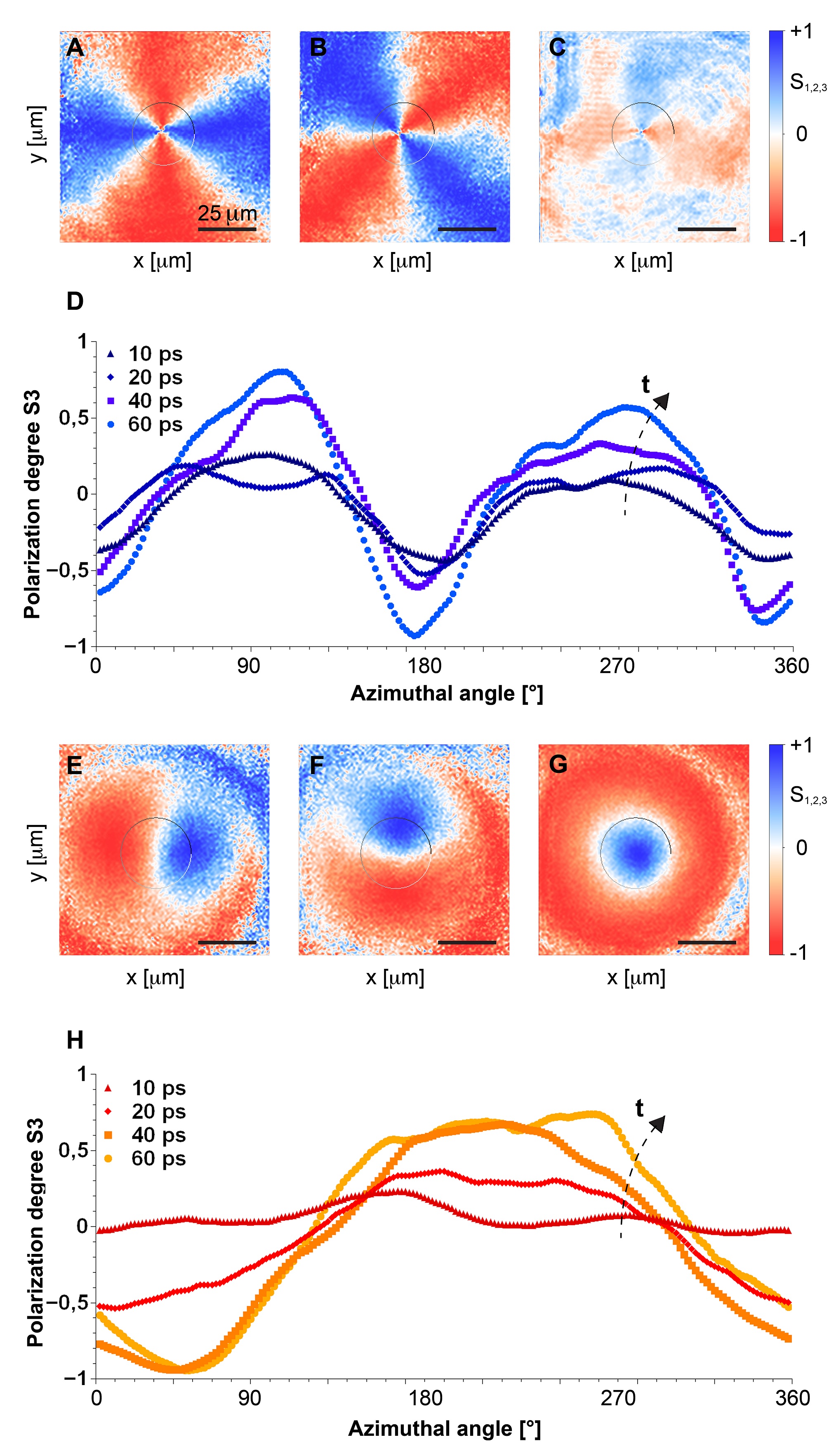}
\caption{Stokes maps. (\textit{A-C}) Polarization maps
  of the polariton hyperspin vortex at the initial time, given as
  a projection onto the three Stokes basis, horizontal-vertical
  ($S_{1}$, \textit{A}) diagonal-antidiagonal ($S_{2}$, \textit{B}) and circular
  right-left ($S_{3}$, \textit{C}). (\textit{D}) Time evolution of the circular degree
  of polarization, plotted as $S_{3}$ angular profiles around the
  vortex core at different times. (\textit{E-G})
  $S_{1}$, $S_{2}$ and $S_{3}$ maps of the lemon-like skyrmion at
  initial time and (\textit{H}) associated time evolution of the spin degree
  profile $S_{3}$. 
  All the profiles have been taken along
  the black/white circle plotted in the real space maps,
which represents the $l$-line of the skyrmion.
} 
\label{fig:FIG2} 
\end{figure}

The most pronounced effect appears in the time evolution of the
  spin degree: the $S_{3}$ component increases during the system
  evolution, becoming almost as large as the linear degree of
  polarization. This effect is clearly visible in Fig.~2\textit{D}, where we
  show the azimuthal profile of $S_{3}$ (taken along the gray circle
in Fig.~2\textit{C}) whose sinusoidal modulation increases with time.
Similar dynamics are observed also when starting with a skyrmion state.
Fig.~2\textit{E-G} shows, for the skyrmion configuration, the associated $S_{1,2,3}$ maps with a lowered
symmetry in the $S_{1}$ and $S_{2}$ linear degrees and a concentric
distribution for $S_{3}$, which changes from $-1$ to $+1$ from the centre
outwards.
We plot the azimuthal profile of $S_{3}$
  along the $l$-line of purely linear polarizations (gray circle in
panel \textit{G}), at different times, in Fig.~2\textit{H} (see also \textcolor{blue}{Movie \hyperref[sec:S1]{S1}}). 
Although the circular degree of polarization
  is approximately zero at the initial time, during time evolution an increasing
  imbalance of right and left spin polarizations develops. Also, in
this skyrmion case, the profile assumes a sinusoidal modulation
growing in amplitude, and rising even larger than in the case of the
polarization vortices. We checked that this effect is not due to a
real-space movement of the whole topological state with respect to
the initial circle. Indeed the phase singularities (vortex cores), that
can be tracked for each spin component possessing a nonzero phase
winding (either two phase singularities for the polarization vortices,
or one for the skyrmions), remain quite stable during the whole dynamics
with just a few $\upmu$m displacement even after 45 ps (see also \textcolor{blue}{Movie \hyperref[sec:S2]{S2}}).

\begin{figure}[htbp]
\centering
\includegraphics[width=0.94\linewidth]{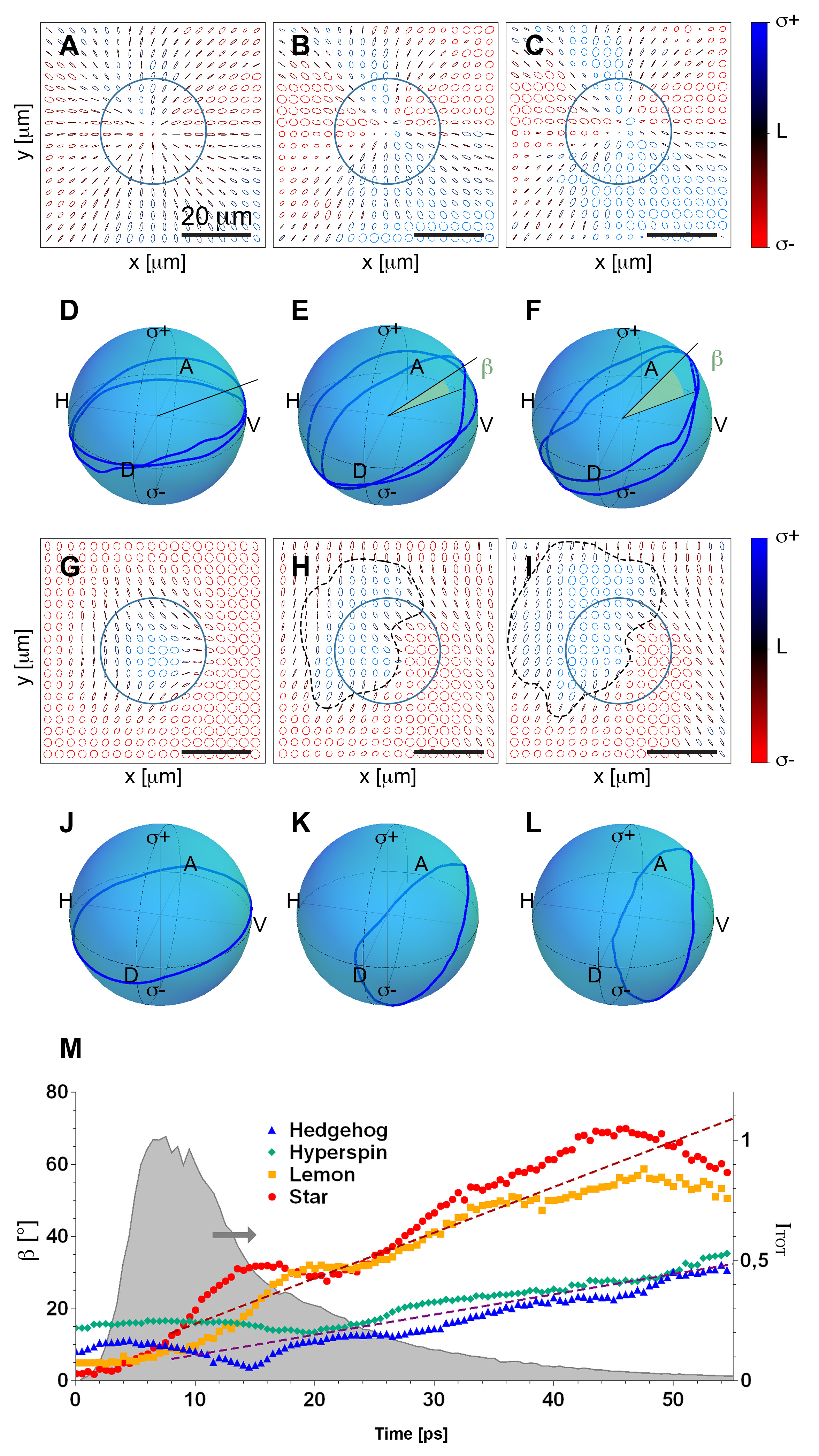}
\caption{Polarization textures. (\textit{A-C})
    Experimental polarization textures of the hedgehog vortex at 20,
    45, 54 ps and (\textit{D-F}) conformal mapping onto the Poincar\'{e} sphere
    of the real-space profile (blue solid circle line in the both kind
    of maps). (\textit{G-I}) Polarization texture of the star-like skyrmion at 20,
    45, 54 ps and (\textit{J-L}) associated plotting in the Poincar\'{e}
    space. The solid blue circle represents the initial boundary line between $\sigma_+$ and $\sigma_-$ spin domains ($l$-line),
while its evolution at later time is reported as a dashed black line.
(\textit{M}) Twisting dynamics, represented by the angle $\beta$
    between the plane containing the single- and double-loops
    around the polarization sphere and the equatorial plane as a
      function of time. The time behavior of the total population
    (grey shaded area) shows that the twisting dynamics are basically
    independent on the polariton density
(straight dashed lines are just a guide for the eye).}
\label{fig:FIG3} 
\end{figure} 

The same effects are observed when starting with a hedgehog vortex and
with a star skyrmion. In particular, in Fig.~3, we plot the full
  polarization vectors in real-space, retrieved from the $S_{1,2,3}$
  maps. The first row shows the polarization vector for the hedgehog
at three different time frames. At the initial time, Fig.~3\textit{A}, the
field pattern follows the classic hedgehog structure schematically
introduced in Fig.~1\textit{D}. The colours used here refer directly to the
degree of spin polarization $S_{3}$. 
In addition, in Fig.~3\textit{D}, we show
the double loop of the pseudospin along the Poincar\'{e} sphere.
The subsequent dynamics are presented as vector maps in Fig.~3\textit{B},\textit{C} (see also \textcolor{blue}{Movie \hyperref[sec:S3]{S3}}). 
On the Poincar\'{e} sphere, Fig.~3\textit{E},\textit{F},  we
observe a clear twist of the plane containing the double loop away
from the equatorial plane, where this effect grows in time (see also \textcolor{blue}{Movie \hyperref[sec:S4]{S4}}). 
The twist angle $\beta$
is directly linked to the maximum degree of circular polarization
assumed by the polariton population, as $\beta = \max\arcsin S_3$. By
sinusoidal fitting of the azimuthal profiles of $S_{1,2,3}$, 
we retrieve the trajectory and twist angle $\beta$ of the
double loop. 

Analogous effects are observed for the star skyrmion, which vector textures are
shown in the third row.  The experimental map at the
initial time, Fig.~3\textit{G}, is very close to the sketch pattern of
Fig.~1\textit{B}. The polarization reshaping at later times, as shown in
Fig.~3\textit{H},\textit{I}, results in an apparent spin transport 
with respect to 
the inner circle, initially containing prevalent positive spin ($\sigma_+$), 
both outwards (from the top-left area)
and inward (to the bottom-right part).
Overall, by considering the sign of the spin, 
the entering negative currents ($\sigma_-$)
contribute to the net outgoing positive spin flux.
We should
  emphasize that the spin transport is decoupled from the mass
transport and from the phase singularity movement.
Yet again, it is possible to clearly follow the dynamics on the
Poincar\'{e} sphere, as in Fig.~3\textit{J-L}, where the single-loop
initially on the equator undergoes a large twist also for this
case. 
A similar reshaping effect has been observed before for the
spontaneous hyperbolic spin vortex generated in non-resonant quasi-cw
condition by Manni \textit{et al.}~\cite{Manni2013}. However, in that
case the vortex was instead pinned by a defect and the
  causes for twisting were ascribed to the interplay between
  the disorder potential and the finite-$k$ TE-TM splitting, relevant due
  to radial flows of polaritons.\\

\noindent \textbf{Theory models and discussion\\}
In order to understand the physical origins of our observations, we
  perform numerical modelling of the system's dynamics using
  two-component open-dissipative Gross-Pitaevskii equations, which
  describe the microcavity photon field and the quantum
  well exciton field coupled to each other.
The excitonic coupling between differently polarized populations
is usually represented by the inter-spin nonlinearity term~\cite{Ferrier2011}, 
although here we are interested in the linear regime.
 Fundamental to the present work are
 the terms directly acting on the photonic fields
(see \textcolor{blue}{\hyperref[sec:Sinfo]{\textit{Supporting Information}}} for the model details), 
as discussed in the following.
The photonic coupling between different polarizations is given by the
finite-$k$ TE-TM splitting term $\chi$ and the $k$-independent
anisotropy splitting $\chi_0$. The former appears due to the
difference of transverse-electric and transverse-magnetic masses of
microcavity modes~\cite{bliokh_spinorbit_2015}, while the
$k$-independent splitting $\chi_0$ between linearly polarized modes
can be present in some samples due to strain
effects~\cite{balili_huge_2010,klopotowski_optical_2006} and
heavy-light hole mixing~\cite{ivchenko_heavy-light_1996} on the
quantum well interfaces. In such cases two linearly polarized waves
with specific polarization directions (say, $x$ and $y$), are subject
to a slight different energy shift, regardless of the direction and
of their wavevector. Hence, also the $k=0$ state can be subject
to a dephasing between the two linear components, and as a result
there could be a precession of an initial polarization state
(different from the linear $x$ and $y$ ones) at each point in space.

\begin{figure}[htbp]
\centering
\includegraphics[width=1.00\linewidth]{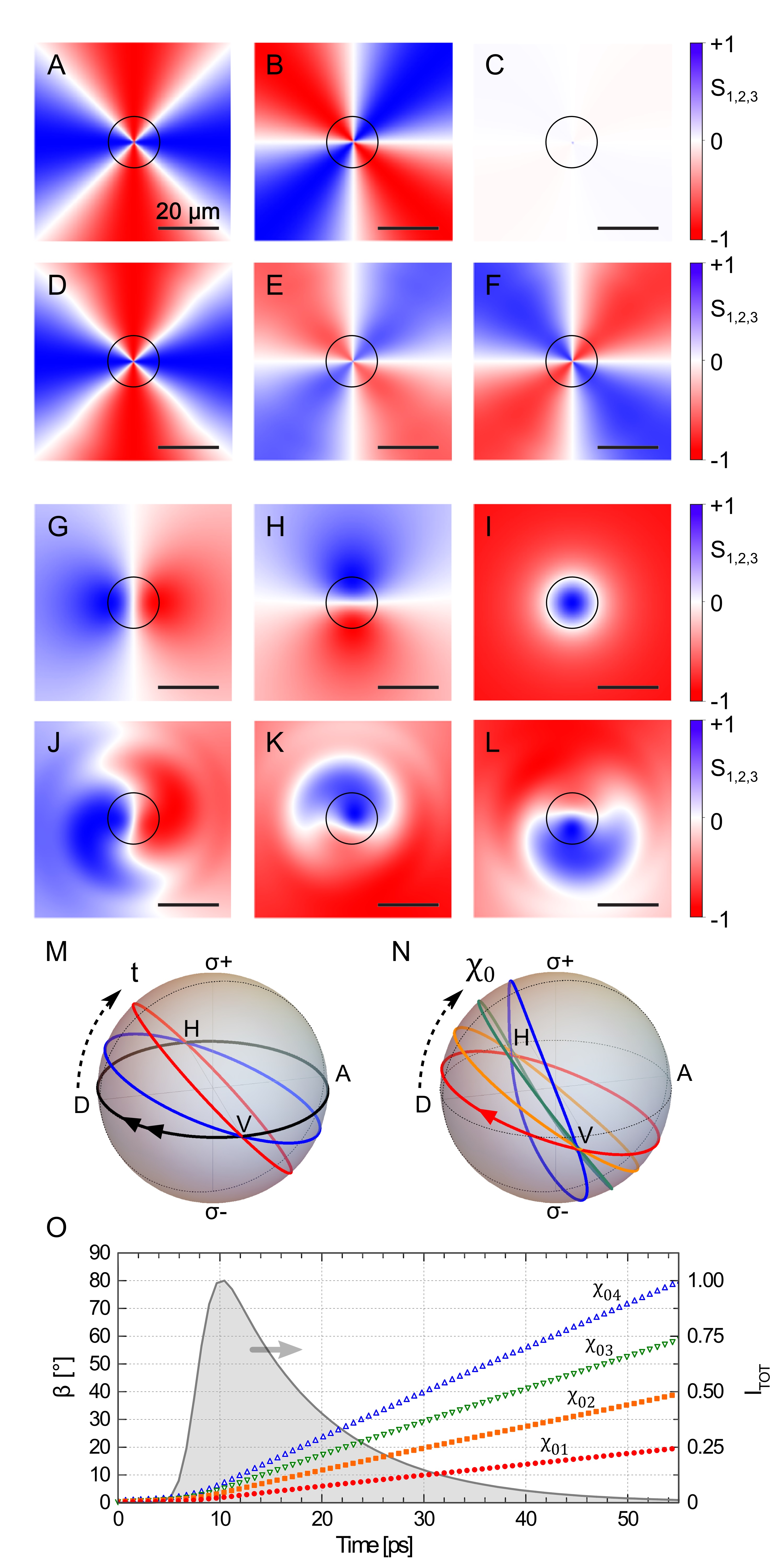}
\caption{Theoretical analysis. (\textit{A-C})
    Polarization maps of the hedgehog vortex from numerical
      simulations at the early stage and (\textit{D-F}) final time
    ($t=67~\text{ps}$). (\textit{G-I}) $S_{1}$, $S_{2}$ and $S_{3}$ maps of the
    star-like skyrmion at initial and (\textit{J-L}) final time
    ($t=67~\text{ps}$). (\textit{M-N}) Twisting of the associated double-loop
    (\textit{M}) and single-loop (\textit{N}) trajectory in the Poincar\'{e} sphere,
    retrieved along the black circle in the real space maps
     (which is the initial $l$-line of the skyrmion). 
    In panel \textit{M} the loops are traced at different times ($t = 6~\text{ps},~34~\text{ps}$ and $67~\text{ps}$) and constant anisotropy ($\chi_0 \equiv \chi_{02}$),
while in \textit{N} the loops are traced at fixed time ($t = 55~\text{ps}$) and different $\chi_0$ values
 ($\chi_{01} = 0.01~\text{meV},~\chi_{02} = 0.02~\text{meV},~\chi_{03} = 0.03~\text{meV},~\chi_{04} = 0.04~\text{meV}$)
    (\textit{O}) Time
    evolution of the twist angle $\beta$ for the same anistropy values used in panel \textit{N}.
The twist rate is constant in time and linear on the $\chi_0$ value,
and independent on the topology state
 (which is, 
each $\beta(t)$ twist curve in the panel \textit{O} 
is the same for all the four states, 
hedgehog and hyperspin, lemon and star skyrmion). 
 }
\label{fig:FIG4} 
\end{figure}

To reproduce the polarization twisting
observed in our experiment we perform different sets of simulations. In the
first set we take the $k$-independent anisotropy splitting $\chi_0$ to
be zero, and in the second set we assume it to be in the 
$\chi_0=0.01~\text{meV}-0.04~\text{meV}$ range, 
estimated on the basis of polarization
resolved photoluminescence. We see no twist effect in the dynamics
when $\chi_0=0$, and instead see a significant twist,
comparable to experimental, in the simulations with
$\chi_0$ inside the said range (Fig.~4).
The initial $S_{1,2,3}$ Stokes maps for the spin vortices are
reported in Fig.~4\textit{A-C}, respectively, for the case of a hedgehog
state. Here the initial degree of circular polarization is
homogeneously null ($S_3$ map of panel \textit{C}). The evolution of
  $S_{1,2,3}$ at later times ($t=67~\text{ps}$), presented in
the second row of Fig.~4\textit{D-F}, shows emergence of a strong circular
  polarization under the action of a $\chi_0 \equiv \chi_{02} = 0.02~\text{meV}$. The $S_3$ map (panel \textit{F}) exhibits a symmetric
  division in four quadrants aligned as those of the $S_2$ parameter
  (panel \textit{E}).  $S_2$ is maintaining the same orientation as in the initial
  state but is decreasing in its intensity, while the $S_1$ is
  essentially unmodified in both orientations and intensity.  This
effect is indeed observed only in the presence of a $k=0$ $xy$
anisotropy, which in the simulations has the specific orientation
along the $x$ and $y$ axis and thus is not affecting the $S_1$
pattern. On the Poincar\'{e} sphere, the space circulation of the
polarization vortices around the cores at the initial time can be
mapped to a double rotation on the sphere lying in the equatorial
plane as in Fig.~4\textit{M}. The effect of the dynamical polarization reshaping
is equivalent to a twist of the geodesics around the $S_1$ axis
towards the circular poles, which grows in time.

Similar effects are observed when starting with a skyrmion. As an
  example in Fig.~4\textit{G-I} we show the star-like state, with their
  associated two-sector symmetry in the linear polarizations. 
  Here the degree of circular polarization is not zero at the initial
  state due to the skyrmion structure, which translates to a vortex
  in one circular polarization and the Gaussian state in the other.
The polarization evolves in time in a similar way to what we have seen
for the skyrmions in the experiment, and is caused by the
mechanism associated with the $\chi_0$ splitting, as described earlier.
Figure~4\textit{J-L} shows the Stokes maps obtained at later time ($t=67~\text{ps}$)
again under the action of a $\chi_0 \equiv \chi_{02}$ anisotropy value.
We also examine the polarization profile along a circle in real space
 taken along the so called $l$-line, marked on
the maps as a black solid circle. This is conformal to a single
loop around the equator of the Poincar\'{e} sphere as shown the
Fig.~4\textit{N}. Here we report the loops at fixed time 
($t = 55~\text{ps}$)
and for different increasing $\chi_0$ values
($\chi_{01} = 0.01~\text{meV},~\chi_{02} = 0.02~\text{meV},~\chi_{03} = 0.03~\text{meV},~\chi_{04} = 0.04~\text{meV}$).
The polarization reshaping with its associated Stokes twist is
once again happening along the $S_1$ axis and it is proportional to the $\chi_0$ strength (see also \textcolor{blue}{Movie \hyperref[sec:S5]{S5}}).
The initially circular symmetry of the spin degree in real space
evolves as well, as seen in Fig.~4\textit{L}. It assumes a distribution which
is somehow complementary to that of the $S_2$ one.\\

\noindent \textbf{Conclusions and Perspectives\\}
In summary, the dephasing of $x$ and $y$ linear polarization
components, leads to a transformation of the diagonal-antidiagonal
degree of polarization into a circular spin degree, for both spin vortices and
skyrmions. By comparing with the experiments, we deduce that the
axis of the $k=0$ splitting anisotropy in our experimental
configurations are oriented along the diagonal and antidiagonal
directions.  The twist speed induced by the $\chi_0$ term in the
simulations is the same for all four considered states. The perfectly
linear trend in time (starting from $t=8~\text{ps}$, that is when the
fluid is left free to evolve after the arrival of the exciting pulse)
shown in Fig.~4\textit{O} demonstrates the effect to be independent of the
instantaneous density of polaritons, which decay according to the
$\tau_{\text{LP}} = 10~\text{ps}$.  
On the other hand the strength of the twist is directly proportional to the anisotropy value,
as demonstrated by looking at the slopes of the $\beta$ curves
 in Fig.~4\textit{O} corresponding to different $\chi_0$. 
We can evaluate a theoretical twist speed of $\sim40^{\circ}(\text{ps}\cdot\text{meV})^{-1}$.
Noticeably here (for the skyrmions), we also
observe an interesting evolution of the $S_1$ pattern in real space
(Fig.~4\textit{J}). There is a sort of rotation of the sectors with some
features of spiralling. This additional effect, i.e., the rotations of
the $S_{1,2,3}$ sectors in real space, is instead associated to the
action of the finite-$k$ TE-TM splitting term $\chi$ in our model~\cite{Cilibrizzi2015}.  We
would like to stress that our simulations clearly confirm that it is
the $xy$ anisotropy, which is the cause for the polarization twisting of
vortex states. On the contrary, the disorder potential term,
produced by the inhomogeneities inside the cavity mirror, 
was not needed to
reproduce the observed dynamics.

From a theoretical point of view, the reshaping of the polarization
field, and more specifically their Stokes twist, can be a convenient
way to define the generalized quantum vortex, where the angle $\beta$
measures the inclination between the plane of polarization rotation
and the equatorial plane in the Poincar\'{e} sphere.  We note that the
concept of generalized quantum vortex can be used to describe the new
type of half-quantum circulation, recently found in a macroscopic ring by Liu \textit{et
  al.}~\cite{Liu2015} under non-resonant pulsed pumping. Namely, this
vortex corresponds to the polarization rotation around a tilted axis on
the pseudospin sphere. 
We can hence define the generalized skyrmion as a
full Poincar\'{e} topology which real-space circuitations are conformal to
a family of single loop curves around an arbitrary axis on the pseudospin sphere (see \textcolor{blue}{Movie \hyperref[sec:S6]{S6}}).  
The same concept
can apply to the spin vortices, whose generalized version maps to a
double loop along an arbitrary great circle of the
Poincar\'{e} sphere.\\

\noindent \textbf{Methods summary\\}
\small{
The experimental polariton device is a typical photonic microcavity (MC)
embedding quantum wells (QW) kept at cryogenic temperature. 
A ps laser pulse tuned on the lower polariton branch energy 
works as the excitation and reference beams.  
Optical vortices and their composition are obtained by
means of a liquid crystal $q$-plate device, waveplates and polarizers.
Space-temporal dynamics are retrieved upon implementing
the off-axis digital holography technique on a custom interferometric setup.
The modelling of the system 
is based on coupled two-component 
open-dissipative Gross-Pitaevskii equations for the MC photons and QW excitons.
Dynamical simulations of the equations are implemented 
on the XMDS2 software framework~\cite{dennis_xmds2_2013}.
For experimental and theoretical details 
 see refs.~\cite{Dominici2015,Dominici2014}, and~\cite{Colas2015} 
and \textcolor{blue}{\hyperref[sec:Sinfo]{\textit{Supporting Information}}}.\\
}

\noindent \textbf{ACKNOWLEDGEMENTS.}
\small{We thank R.~Houdr\'{e} for the 
microcavity sample and L.~Marrucci and B.~Piccirillo for the $q$-plate
devices. 
This work was supported
by European Research Council POLAFLOW Grant 308136, Italian Ministero
dell'Istruzione dell'Universit\'{a} e della Ricerca project ''Beyond Nano'',
Engineering and Physical Sciences Research Council Grants EP/I028900/2 and
EP/K003623/2 (to M.H.S.), and Consejo Nacional de Ciencia y Tecnolog\'{i}a
Grant 251808 (to Y.G.R.).
}




%

\cleardoublepage
\pagebreak
\clearpage 

\setcounter{equation}{0}
\setcounter{figure}{0}
\setcounter{table}{0}
\setcounter{page}{1}
\makeatletter
\renewcommand{\theequation}{S\arabic{equation}}

\renewcommand{\figurename}{\textbf{Movie}}
\renewcommand{\thefigure}{\textbf{S\arabic{figure}}}

\renewcommand\thepage{S\arabic{page}}

\phantomsection{}
\label{sec:Sinfo}
\noindent
\textbf{\LARGE Supporting Information}\\

\normalsize

\noindent \textbf{\Large Twist of generalized skyrmions and spin vortices in a polariton superfluid\\}

\noindent \textbf{SI Text\\}

\noindent \textbf{Experimental methods.}
The experimental polariton device is an AlGaAs 2$\lambda$ microcavity
with three 8 nm In$_{0.04}$Ga$_{0.96}$As quantum wells.  All the
experiments shown here are performed at a temperature of 10~K in a
region of the sample clean from defects.  The excitation beam is a
4.0~ps Gaussian laser pulse with a repetition rate of 80 MHz
selectively tuned on the lower polariton branch energy.  Its intensity
is adjusted so that to keep the resonantly excited fluid in a
linear regime during the whole dynamics.  In order to obtain the four
different initial topological patterns (as reported in the table of
Fig.~1\textit{M}) we used a combination of impinging polarization, electrical
tuning of the $q$-plate and waveplates as described below.  In the
case of the spin vortices the pulse is linearly polarized and the
tuning of the $q$-plate is complete (100\%). This allows to directly
obtain a hedgehog pattern at the exit.  Upon insertion of a
half-wave plate (HWP) after the $q$-plate, we locally rotate the
linear vectors of such pattern, obtaining the hyperspin topology.  In
the case of the skyrmion, we send the pulse with a circular
polarization onto the $q$-plate which is now partially tuned (50\%).
This results in an outcoming lemon skyrmion, which can be rotated by means
of an HWP into its conjugated state, the star skyrmion.

On the detection side, to obtain polarization-resolved imaging, a
waveplate and a linear polarizer are inserted before the charge
coupled device.  Upon using a HWP before the polarizer it is
possible to resolve every direction of the linear polarization ($H$,
$V$, $D$ and $A$), while replacing the HWP with a quarter-wave plate
is possible to map the circular polarizations ($R$ and $L$). In
this way we perform six dynamical sequences for each initial topology,
from which it is possible to extract each independent degree of
polarization. The three Stokes parameters are effectively derived as
$S_1=\frac{I_H-I_V}{I_H+I_V}$, $S_2=\frac{I_D-I_A}{I_D+I_A}$ and
$S_3=\frac{I_{R}-I_{L}}{I_{R}+I_{L}}$, where the intensities are a
function of both time and space [e.g., $I_H(x,y,t)$]. We checked that
the total intensity in each of the three basis is the same at
  each point in space and time,
$(I_H+I_V)(x,y,t)=(I_D+I_A)(x,y,t)=(I_R+I_L)(x,y,t)$.  In other terms,
that there is no significant depolarization and the six measurements are
consistent with each other.

To obtain the time dynamics, the emission profiles are made to
interfere with a delayed expanded reference beam carrying homogeneous
density and phase profiles.  Such technique is known as off-axis
digital holography and relies on the use of Fast Fourier Transform (FFT)
to filter only the information associated with the simultaneity
between the emission and the delayed reference pulse.  In this way it
is possible to study the dynamics of the polariton fluid, by obtaining
the 2D real space snapshots of both the emission amplitude and phase,
at a given time frame set by the delay.  Each final snapshot results
from thousands of repeated events, whose stability is based on the
repeatability of the dynamics (with respect to the physics of the
polaritons) and on the acquisition speed of each single interferogram
(with respect to the experimental setup).  Despite the fact that here
we mostly used intensity features in each of the six pseudospin
vectors, to study the polarization degree distribution and evolution
it is also possible to look at the phase maps to devise the phase
singularities at the cores of the vortex states (which here we
did to check  their stability in time).  Additional details on
the technique and the sample can be found in
refs.~22, 30, and 33.\\

\noindent \textbf{Theory models.}
In order to understand the physical origins of our observations, we
  perform numerical modelling of the system's dynamics using
  two-component open-dissipative Gross-Pitaevskii equations, which
  describe the microcavity photon field $\phi_\pm$ and the quantum
  well exciton field $\psi_\pm$ coupled to each other:
\begin{eqnarray}
  &  i\hbar\frac{\partial\phi_{\pm}}{\partial t} =  
  \left(-\frac{\hbar^{2}}{2m_{\phi}}\nabla^{2}-i\frac{\hbar}{2\tau_{\phi}}\right)\phi_{\pm}
  +\frac{\hbar\Omega_{\text{R}}}{2}\psi_{\pm}\nonumber\\ 
  & + \chi\left(\frac{\partial}{\partial x}\mp
    i\frac{\partial}{\partial y}\right)^{2}\phi_{\mp} + \frac{1}{2}\chi_0\phi_{\mp} + D\phi_{\pm} + F_{\pm}\label{eq:total} \\
  & i\hbar\frac{\partial\psi_{\pm}}{\partial t} 
  =\left(-\frac{\hbar^{2}}{2m_{\psi}}\nabla^{2}-i\frac{\hbar}{2\tau_{\psi}}\right)\psi_{\pm}
  +\frac{\hbar\Omega_{\text{R}}}{2}\phi_{\pm}\nonumber \\ 
  & + \alpha_1|\psi_{\pm}|^{2}\psi_{\pm} + \alpha_2|\psi_{\mp}|^{2}\psi_{\pm}.\nonumber  
\end{eqnarray}
%


Here the upper lines of both equations represent analogous terms for
the two fields, which are the kinetic energy, the decay time and the
Rabi coupling strength between photons and excitons, respectively.  In
practical terms, excitons have an effective mass $m_{\psi}$ of 4-5
orders of magnitude greater than that of the microcavity photons
$m_{\phi}$, resulting in their kinetic energy being negligible. The
exciton and photon lifetimes are $\tau_{\psi}=1000~\text{ps}$ and
$\tau_{\phi}=5~\text{ps}$, respectively, giving the lower
polariton lifetime of $\tau_{\text{LP}}\approx10~\text{ps}$ at zero detuning
and $k=0$. The Rabi coupling is 
$\hbar\Omega_{\text{R}}=5.3~\text{meV}$. Selective
  excitation of the LPB can be obtained by using ps pulses with less
than $1~\text{meV}$ energy width, tuned on the lower polariton mode as
in the experiments. The bottom lines in both equations represent the
specific terms acting on the two fields. 
The exciton-exciton interaction strengths used in the simulations are
  $\alpha_1=+2~\upmu\text{eV}\cdot\upmu\text{m}^{2}$ for the intra-spin
  nonlinearities and
  $\alpha_2=-0.2~\upmu\text{eV}\cdot\upmu\text{m}^{2}$ for the inter-spin
  ones~(34).  
However, in the present work we are interested in the linear regime and
specifically in the terms
directly acting on the photonic fields, as discussed in the following.

The photonic linear coupling between different polarizations is given by the
finite-$k$ TE-TM splitting term $\chi$ and the $k$-independent
anisotropy splitting $\chi_0$.  The former appears due to the
difference of transverse-electric and transverse-magnetic masses of
microcavity modes~(35) as
$\chi=\frac{\hbar^{2}}{4}(\frac{1}{m_{\phi}^{\text{TE}}}-\frac{1}{m_{\phi}^{\text{TM}}})$,
where the two effective masses imbalance is assumed
$m_{\phi}^{\text{TE}}/m_{\phi}^{\text{TM}} = 0.95$ in our case.  
The $k$-independent splitting $\chi_0$ between linearly polarized modes,
which is due to strain
effects~(36, 37) and
heavy-light hole mixing~(38) on the
quantum well interfaces, results in a different energy shift
between the relevant linear polarized modes,
and in the accumulation of a relative phase.
In our simulations, we assumed $xy$ directions 
for the anisotropy axis, 
and used 4 different $\chi_0$ values
($\chi_{01} = 0.01~\text{meV},~\chi_{02} = 0.02~\text{meV},~\chi_{03} = 0.03~\text{meV},~\chi_{04} = 0.04~\text{meV}$).
 
The disorder potential term $D(x,y)$,
produced by the inhomogeneities inside the cavity mirror, 
which is reported here only for completeness, was not used and not needed to
reproduce the observed dynamics.
Finally, the initial laser pulse is described as a pulsed
Laguerre-Gauss $F_{\pm}$:
\[
F_{\pm}(\mathbf{r})=f_{\pm}r^{|l_{\pm}|}e^{-\frac{1}{2}\frac{r^{2}}{\sigma_{r}^{2}}}e^{il_{\pm}\theta}e^{-\frac{1}{2}\frac{\left(t-t_{\text{0}}\right)^{2}}{\sigma_{t}^{2}}}e^{i(\mathbf{k}_{p}\cdot\mathbf{r}-\omega_{p}t)} 
\]
with a winding number of the vortex state in the $\sigma_{\pm}$
component represented by $l_{\pm}$, and a strength $f$ that reproduces
the total number of output photons. The parameters of the initial
state are chosen to reproduce the
experimental specifics ($\sigma_{\text{r}}$ and $\sigma_{t}$ resulting
in space and time FWHM equal to $30~\upmu\text{m}$ and $4~\text{ps}$,
respectively). The initial state is centered on the LPB mode at 836 nm
and at $\mathbf{k}_{p}=0$.\\ 

\noindent \textbf{Computational methods.}
The dynamics of equation~S1 is simulated using the XMDS2 software
framework~(40). We employed adaptive step-size algorithm based on
fourth and fifth order ``embedded Runge-Kutta'' (ARK45) method with
periodic boundary conditions. This algorithm was also tested against
eighth and ninth order (ARK89) of ``embedded Runge-Kutta'' method. 
The periodic boundary condition is an artefact of using FFT in order to efficiently 
switch between the real space to compute the potential energy and the momentum space to evaluate the kinetic energy. 
This method ensures very fast computation of each time step.
In order to ensure that all flux leaving the system is not coming back
from the other side due to the periodic boundary conditions, we
implemented additional circular/ring absorbing boundary conditions,
with the depth and the width carefully adjusted to the geometry of
current experiments. We solve the equations on a 2D finite grid of $N
\times N = 1024 \times 1024$ points and lattice spacing $l=0.54~\upmu\text{m}$ 
in a box of $L\times L=556 \times 556~\upmu\text{m}^2$. 
The large size of the simulation box ensures that polariton
density drops practically to zero at the boundary. However, in all the maps  
we plot the physically relevant central region only, where the
density of polaritons is still significant.\\

\vspace{2cm}

\noindent \textbf{SI Movies}

\phantomsection{}
\label{sec:S1}

\begin{figure}[H]
\centering
\includegraphics[width=1.0\linewidth]{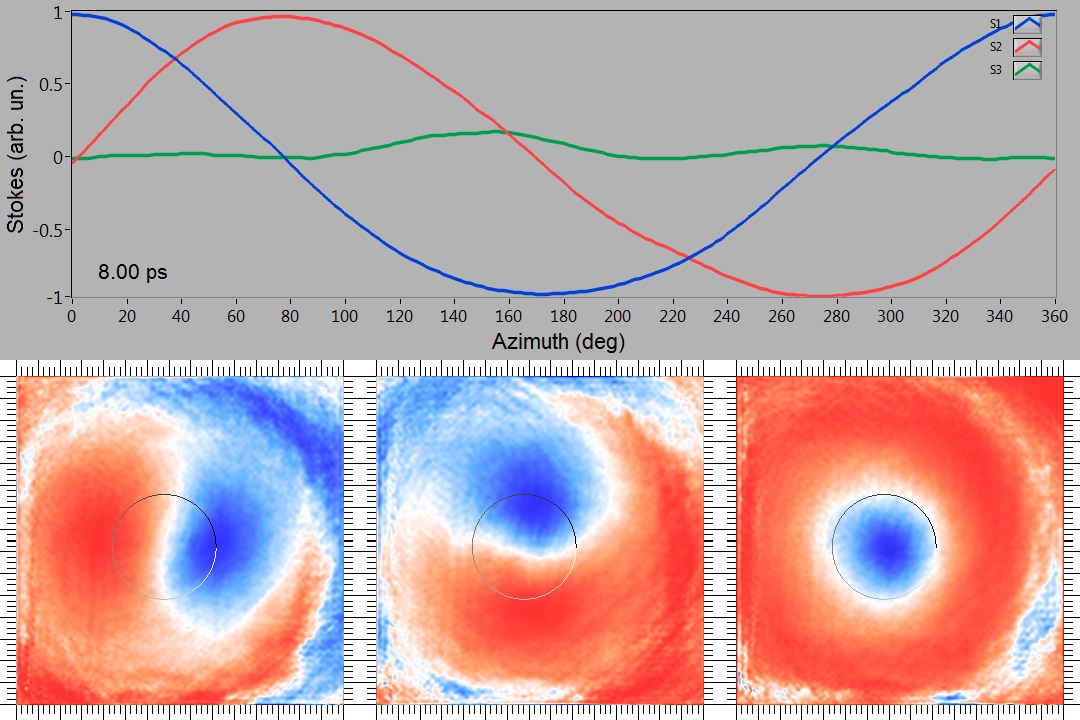}
\caption{
Lemon skyrmion, dynamics of the $S_{1,2,3}$ Stokes maps on a $90 \times 90~\upmu\text{m}^2$ wide area and with time step of $0.5~\text{ps}$. The top line shows the associated azimuthal profiles computed along the real-space circle represented in the maps, as in the case of Fig.~2\textit{E-H}.
The initially flat profile of the $S_3$ circular degree of polarization grows in time assuming a sinusoidal modulation as large as the other two Stokes parameters.
\href{http://movie-usa.glencoesoftware.com/video/10.1073/pnas.1610123114/video-1}{\textcolor{blue}{Movie S1}} 
 }
\label{fig:FIGS1} 
\end{figure}

\phantomsection{}
\label{sec:S2}

\begin{figure}[H]
\centering
\includegraphics[width=1.0\linewidth]{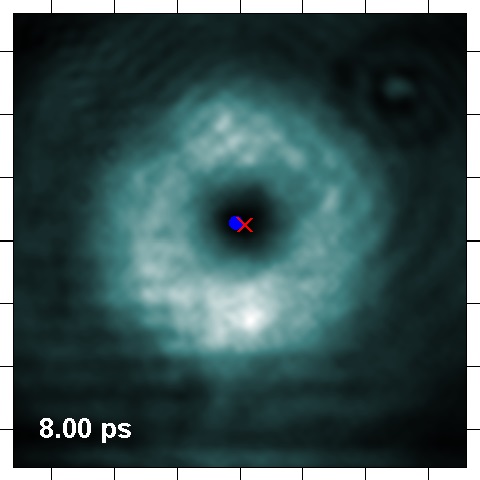}
\caption{
Hedgehog vortex, dynamics of the total polariton density distribution on a $72 \times 72~\upmu\text{m}^2$ area.
The associated positions of the counter-winding phase singularities in the two spin components are represented too, to show their fundamental stability over the whole dynamics 
(blue dot corresponds to the $\sigma_+$ and red cross to the $\sigma_-$ spin population).
\href{http://movie-usa.glencoesoftware.com/video/10.1073/pnas.1610123114/video-2}{\textcolor{blue}{Movie S2}}
 }
\label{fig:FIGS2} 
\end{figure}

\newpage

\phantomsection{}
\label{sec:S3}

\begin{figure}[H]
\centering
\includegraphics[width=1\linewidth]{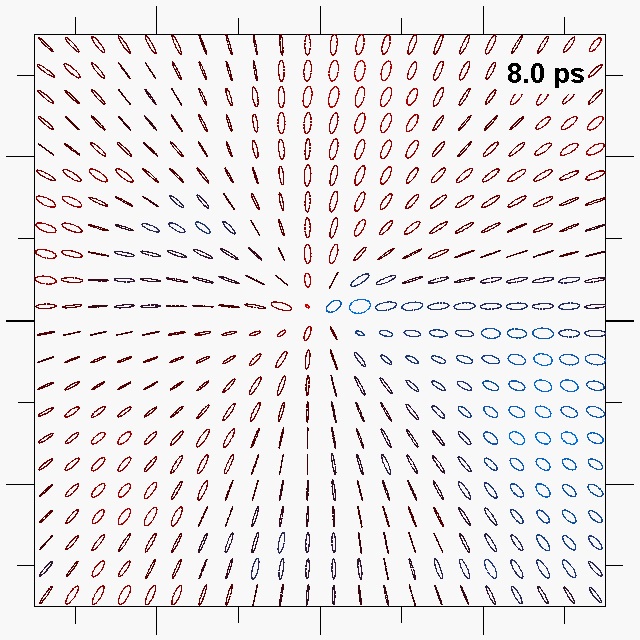}
\caption{
Representation of the polarization texture of the hedgehog vortex and of its reshaping dynamics over a $70 \times 70~\upmu\text{m}^2$ wide area. 
The initially radial field evolves in time assuming an elliptical polarization degree with a four-sectors configuration.
\href{http://movie-usa.glencoesoftware.com/video/10.1073/pnas.1610123114/video-3}{\textcolor{blue}{Movie S3}}
 }
\label{fig:FIGS3} 
\end{figure}

\phantomsection{}
\label{sec:S4}

\begin{figure}[H]
\centering
\includegraphics[width=1.05\linewidth]{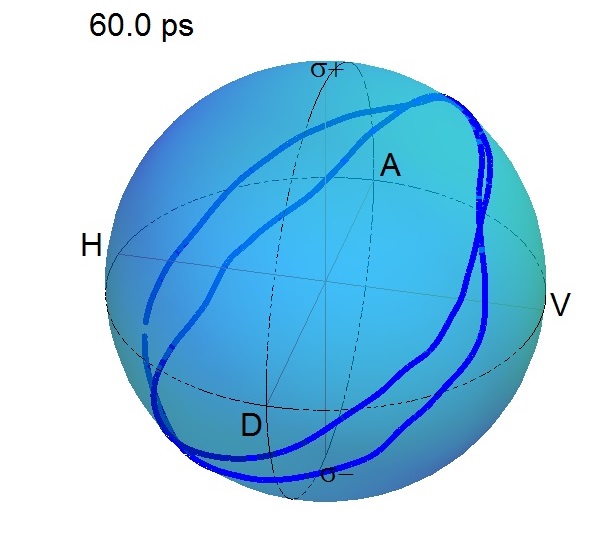}
\caption{
The dynamics of the hedgehog vortex represented by means of its mapping onto the Poincar\'{e} sphere.
The action of $xy$ anisotropy produces a twist of the pseudospin double-loop plane towards the poles of the sphere.
\href{http://movie-usa.glencoesoftware.com/video/10.1073/pnas.1610123114/video-4}{\textcolor{blue}{Movie S4}}
 }
\label{fig:FIGS4} 
\end{figure}

\phantomsection{}
\label{sec:S5}

\begin{figure}[H]
\centering
\includegraphics[width=0.80\linewidth]{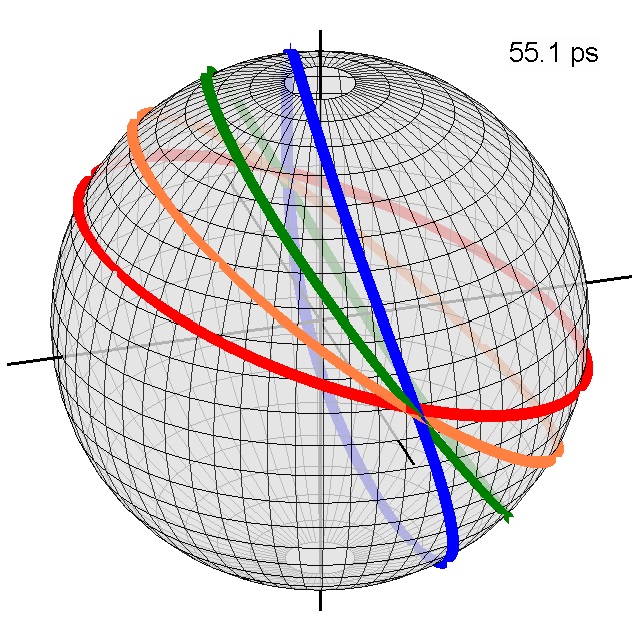}
\caption{
Numerical model dynamics of the lemon skyrmion twist, plotted on the Poincar\'{e} sphere.
Each twisting loop corresponds to a different value of the anisotropy
($\chi_{01},~\chi_{02},~\chi_{03},~\chi_{04}$)
consistently to what reported in Fig.~4\textit{N},\textit{O}.
\href{http://movie-usa.glencoesoftware.com/video/10.1073/pnas.1610123114/video-5}{\textcolor{blue}{Movie S5}}
 }
\label{fig:FIGS5} 
\end{figure}

\phantomsection{}
\label{sec:S6}

\begin{figure}[H]
\centering
\includegraphics[width=1.0\linewidth]{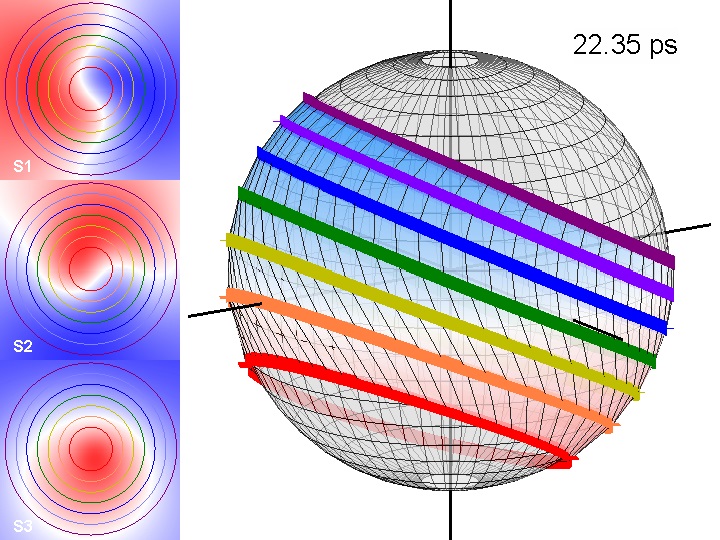}
\caption{
Numerical star skyrmion twist for a value of the anisotropy $\chi_{04}$. The $S_{1,2,3}$ Stokes degree of polarizations are plotted in the left panels over a $60 \times 60~\upmu\text{m}^2$ area. Several concentric circles in real space (each represented with a different color) are conformally mapped onto the Poincar\'{e} sphere on the right panel (the $l$-line corresponds to the green color curve). The dynamical twist is similar for each circle, highlighting the generalized skyrmion features. At around $65~\text{ps}$ the twist angle has reached $90^{\circ}$ and the $S_2$ and $S_3$ distributions in real space have roughly swapped with respect to initial time. Upon longer time the twist leads to a complete reversal of the initial star skyrmion into its conjugate state, the lemon skyrmion. \href{http://movie-usa.glencoesoftware.com/video/10.1073/pnas.1610123114/video-6}{\textcolor{blue}{Movie S6}}
 }
\label{fig:FIGS6} 
\end{figure}

\end{document}